\documentclass[12pt]{article}

\usepackage[body={17.5cm, 22.5cm},right=2cm]{geometry}

\usepackage{color}
\usepackage{graphicx}
\usepackage{hyperref}
\usepackage{epsf}
\usepackage{graphicx,epsfig}
\usepackage[skip=2pt,font=small]{caption}

\usepackage{float}

\usepackage{amsmath}
\usepackage{amssymb}
\usepackage{epsfig}
\usepackage{cite}
\usepackage{color,colordvi}
\newcommand{\be}{\begin{eqnarray}}
\newcommand{\ee}{\end{eqnarray}}
\newcommand{\bi}{\begin{itemize}}
\newcommand{\ei}{\end{itemize}}

\newcounter{hran}

\def\MSbar{\relax\ifmmode\overline{\rm MS}\else{$\overline{\rm MS}${ }}\fi}

\def\d{\rm d}

\def\d{{\rm d}}

\def\vx{\vec{x}}

 \def\vx{\vec{ x}} 
\def\vk{\vec{k}}

\def\D{\Delta}

\def\n3{\nu_3}

\def\n1{\nu_1}

\def\simlt{\stackrel{<}{{}_\sim}}

\begin{document}

\numberwithin{equation}{section}
\thispagestyle{empty}
\vspace{5mm}
\vspace{0.5cm}
\begin{center}

\def\thefootnote{\fnsymbol{footnote}}

{\Large \bf 
High Energy Physics Signatures
from  Inflation\\
and Conformal  Symmetry of de Sitter 
\vspace{0.25cm}
\\
\vspace{0.25cm}	
}
\vspace{1.5cm}
{\large  
A. Kehagias$^{a,b}$ and  A. Riotto$^{b}$
}
\\[0.5cm]

\vspace{.3cm}
{\normalsize {\it  $^{a}$ Physics Division, National Technical University of Athens, \\15780 Zografou Campus, Athens, Greece}}\\

\vspace{.3cm}
{\normalsize { \it $^{b}$ Department of Theoretical Physics and Center for Astroparticle Physics (CAP)\\ 24 quai E. Ansermet, CH-1211 Geneva 4, Switzerland}}\\

\vspace{.3cm}


\end{center}

\vspace{2cm}

\hrule \vspace{0.3cm}
{\small  \noindent \textbf{Abstract} \\[0.3cm]
\noindent 
During inflation, the geometry of spacetime is described by a  (quasi-)de Sitter phase. Inflationary observables are determined by the underlying (softly broken) de Sitter isometry group $SO(1,4)$ which acts 
like a  conformal group  on $\mathbb{R}^3$:  when the fluctuations are on 
super-Hubble scales, the correlators of the scalar fields  are constrained by conformal invariance.    
Heavy fields with mass $m$ larger than the Hubble rate $H$ correspond to operators with imaginary dimensions in the dual Euclidean three-dimensional conformal field theory. By making use of the dS/CFT correspondence we show that, 
besides the  Boltzmann suppression expected from  the thermal properties of de Sitter space,
the generic  effect of heavy fields  in the inflationary correlators of the light fields is to   introduce  
 power-law suppressed corrections of the form
 ${\cal O}(H^2/m^2)$. This can be seen, for instance, at the level of the four-point correlator for which we provide the correction   due to a  massive scalar field  exchange. 
\vspace{0.5cm}  \hrule
\vskip 1cm

\def\thefootnote{\arabic{footnote}}
\setcounter{footnote}{0}

\newpage
\baselineskip= 18pt

\section{Introduction}
Accessing the massive states  of a generic high energy theory requires very energetic phenomena to occur. For instance, at the LHC
one may hope to produce  dark matter particles, which are believed to account for about 30\% of the energy content of the universe, by 
colliding highly energetic protons and thus producing these new weakly interacting particles. In this sense,  cosmological inflation \cite{lrreview} offers a unique opportunity to access very energetic states for two main reasons: the energy scale of inflation (which we identify here with the Hubble rate $H$)  may be very high and all massive states are quantum-mechanically generated (to some level) during inflation due to the fact that the background is time-dependent. 

The inflationary observables are $N$-point correlators of the
curvature perturbations and they are currently (or soon will  be)  investigated  both in the cosmic microwave background radiation
anisotropies and in the large scale structure of the universe. The hope is that these measurements are of such high accuracy to be able to find  traces of the very UV  states of the theory responsible for inflation. One might think, for instance, that the four-point correlators generated during inflation might be written in terms of scattering amplitudes  whose poles and  residues are  determined by all states of the theory, including the very massive ones. This would give the possibility of reconstructing the full theory.
 In this respect, the analogy within the Anti de Sitter/Conformal Field Theory (AdS/CFT) correspondence
\cite{mal} is useful:   the contribution to a four-point function from the exchange of
a bulk scalar field of an arbitrary mass in AdS can be identified with the 
the CFT correlators of the boundary theory
for the   case of  scalar
intermediate states \cite{tse}.

During inflation, the geometry of spacetime is described by a  quasi-de Sitter phase, and one may expect that certain cosmological observables should be determined by the symmetries  of the underlying (softly broken) de Sitter isometries.  
This is the basic underlying   idea of the  so-called  de Sitter/CFT correspondence \cite{strominger0,witten0}.  Since the  de Sitter isometry group $SO(1,4)$ acts  like a  conformal group  on $\mathbb{R}^3$ when the fluctuations are on 
super-Hubble scales, the correlators of the scalar fields  are constrained by conformal invariance    \cite{antoniadis,pimentel,creminelli,noreja,KR1,KR2,KR3,skenderis1,skenderis2,triv}. Unfortunately, there is no notion
of $S$-matrix in de Sitter spacetime (essentially one may not define asymptotic states on super-Hubble scales) \cite{witten0} and therefore reinterpreting the super-Hubble CFT correlators as bulk scatterings is not possible. Nevertheless, one may ask a more modest question: is it possible to use the dS/CFT correspondence to characterize the signatures of the
massive fields in the inflationary observables within some level of  generality? In this paper we take a first step towards the answer of such a question. 

The representations of the de Sitter group $SO(1, 4)$ are devided  into three
distinct series \cite{Dixmie,Vilenkin}, the principal series with masses 
$m^2 \geq 9H^2/4$, the complementary series with
masses in the range $0 < m^2 < 9H^2/4$ and the discrete series. 
Fields in the complementary series are light fields, producing an almost scale invariant power spectrum of scalar perturbations. On the other hand, fields in the principal series are heavy fields and they  correspond to operators with imaginary dimensions in the dual three-dimensional conformal field theory. We will show that 
  the dS/CFT correspondence allows to conclude that states of mass $m\gg H$ appear in the
correlators  in a  power-law suppressed way, that is ${\cal O}(H^2/m^2)$. More in particular, we will provide  the correction to  the four-point correlator
due to the exchange of a  massive scalar field in the bulk and show the it is   power-law suppressed.
 Boltzmann suppressions ${\cal O}[{\rm exp}(-m/H)]$,
 which are expected as  a result of the thermal properties of de Sitter space also appear. 
 
 We should at this point stress that
 our findings ignore other possible effects of heavy fields onto the light field correlators coming from the breaking
 of adiabaticity.  The latter introduces   an extra time scale beside the Hubble time and  might allow exciting  the heavy states
\cite{heavy}. From the CFT point of view, this extra time scale would break the three-dimensional conformal symmetry and therefore this effect may not be captured by dS/CFT correspondence.

The paper is organized as follows.
 In section 2 we recall some known facts about de Sitter space. In  section 3 we discuss the two-point correlators  
 and in section 4 the four-point correlators  for boundary operators and their corrections. We draw some conclusions in section 5. In  Appendix A we recall some facts for the factorization of the Appel's hypergeometric function.

\section{The de Sitter space}
This section contains a short summary of the standard properties of de Sitter spaces. The expert reader may skip it.

The $(n+1)$-dimensional de Sitter spacetime $\d S_{n+1}$ of constant radius $H^{-1}$ \cite{SSV} is described by the hyperboloid 
\be
\eta_{AB}X^AX^B=-X_0^2+\sum_{i=1}^nX_i^2+X_{n+1}^2=\frac{1}{H^2} ~~~~( i=1,\ldots,n),  \label{hyper}
\ee
embedded in  $(n+2)$-dimensional Minkowski spacetime $\mathbb{M}^{1,n+1}$  with coordinates $X^A$ ($A,B,...=0,1,...,n+1$)
and flat metric  $\eta_{AB}=\rm{diag}\,(-1,1,\ldots,1)$.
It can also be described  as the coset 
\begin{eqnarray}
\d S_{n+1}=\frac{SO(n+1,1)}{SO(n,1)},
\end{eqnarray}
where the $SO(n+1,1)$ isometry group of $\d S_{n+1}$ is manifest. The $SO(n+1,1)$ acts linearly in  $\mathbb{M}^{1,n+1}$  with natural generators
\begin{eqnarray}
J_{AB}=X_A\partial_B-X_B\partial_A.
\end{eqnarray}
They obey the $SO(n+1,1)$ algebra
\begin{eqnarray}
[J_{AB},J_{CD}]= \eta_{AD} J_{BC}-\eta_{AC} J_{BD}+
\eta_{BC} J_{AD}- \eta_{BD} J_{AC}.
\end{eqnarray}
It is easy to  verify that the generators 
\begin{eqnarray}
J_{ij},~~~P_0=J_{0,n+1},~~~\Pi_i^{\pm}=J_{i,n+1}\mp J_{i0}
\end{eqnarray} 
satisfy the conformal algebra in $n+1$-dimensions.

The $SO(n+1,1)$ invariant metric on the de Sitter hyperboloid (\ref{hyper}) can be written after an appropriate coordinate system is defined. 
The de Sitter metric is then the induced metric on the hyperboloid from 
the $(n+2)$-dimensional ambient Minkowski spacetime
\be
{\rm d} s_{n+1}^2=\eta_{AB}  \d X^A \d X^B.  \label{metric5}
\ee
 The global coordinates $(\tau,n^I)$ ($I=1,\ldots, n+1)$ 
 cover the entire de Sitter hyperboloid (\ref{hyper}) and  are defined by the parametrization
\begin{eqnarray}
X^0&=&H^{-1} \sinh \tau\,\,\,\,\,\, (-\infty<\tau<\infty),\nonumber \\
X^I&=&H^{-1}\cosh \tau\,  n^I,~~~~  n^In^I\delta_{IJ}=1.
\end{eqnarray}
The metric on the $\d S_{n+1}$ in these coordinates is written as 
\begin{eqnarray}
\d s^2=H^{-2}\left(-\d\tau^2+\cosh^2\tau \d\Omega_{n}^2\right),
\end{eqnarray}
where $\d\Omega_n^2$ is the metric on the unit $S^n$. 
In fact, the global de Sitter coordinates are actually spherical coordinates on the Euclidean sphere $S^{n+1}$ after 
the  analytic continuation of the azimuthal coordinate $\theta$ to $\theta=\pi/2+i \tau$. 

There are also 
 coordinates which cover only part of the de Sitter hyperboloid.  Such coordinates are defined in the Poincar\'e patch for $\eta<0$ and $x^i\in \mathbb{R}^{n}$ by the embedding 
\begin{eqnarray}
 X^0&=&\frac{H^{-2}-\eta^2+|\vec{x}|^2}{-2\eta},\nonumber \\
 X^i&=&\frac{x^i}{-H\eta},\nonumber \\
 X^{n+1}&=&\frac{H^{-2}+\eta^2-|\vec{x}|^2}{-2\eta}.
 \end{eqnarray} 
 The Poincar\'e patch coordinates  $(\eta,\vec{x})$ cover half of the de Sitter hyperboloid, since 
 \begin{eqnarray}
 X^0+X^{n+1}=\frac{1}{-H^2\eta}>0.
 \end{eqnarray}
In fact, the Poincar\'e patch covers  the physical region within the intrinsic horizon $\eta^2>\vx^2$. 
The metric in these coordinates has the familiar expression of the 
$n+1$-dimensional de Sitter metric with conformal time $\eta$ and scale factor $a(\eta)=-1/H\eta$
\be
 {\rm d}s^2=\frac{1}{H^2\eta^2}\left(-{\rm d}\eta^2+{\rm d} \vx^2\right).
\ee 

\begin{figure}[H]
    \centering
    \includegraphics[width=0.3\textwidth]{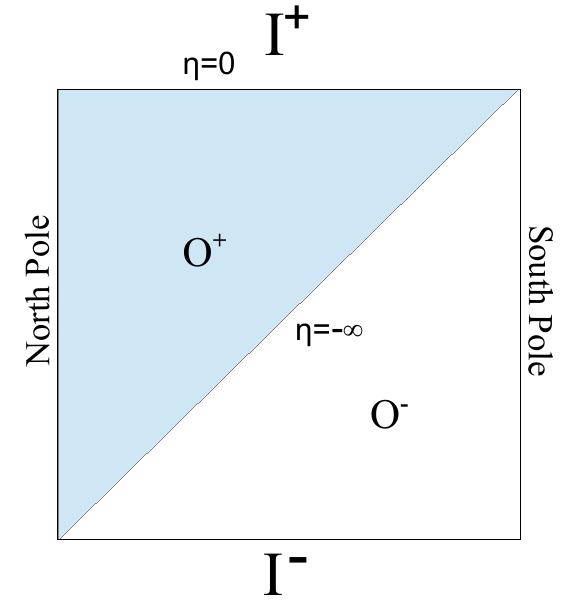}
    \caption{Penrose diagram for de Sitter space.} \label{fig1}
\end{figure}
\noindent
 The Penrose diagram of de Sitter space is dictated in Fig. \ref{fig1}. The $\eta=-\infty$ line divides the spacetime into two regions $O^+$ and $O^-$ and the Poincar\'e patch $|\vec{x}|>\infty, ~-\infty<\eta<0$ covers only $O^+$, the causal future of the observer at the north pole. There are two asymptotic regions $I^+$ and $I^-$, only one of which ($I^+$) is accessible to the observer at the north pole.    

We should note that the Poincar\'e patch is not preserved under the finite action of the isometry group $SO(n+1,1)$. Nevertheless, it is preserved under its infinitesimal action generated by the Killing vectors $J_{AB}$, which in the Poincar\'e patch are written as 
\begin{eqnarray}
&&J_{n+1,i}-J_{0,i}=H^{-1} \partial_i,~~~J_{0i}+J_{n+1,i}=H\left(2x^i\eta\partial_\eta+2 x^ix^j\partial_j-(-\eta^2+x^ix_j)\partial_i\right),\nonumber \\
&& J_{0,n+1}=\eta\partial_\eta+x^i\partial_i, ~~~J_{ij}=x_i\partial_j-x_j\partial_i.
\end{eqnarray}
It is easy to see that at the boundary $I^+$, where $\eta=0$, the Killing vectors turns out to be 
\begin{eqnarray}
&&P_i=H^{-1} \partial_i,~~~K_i=H\left(2 x^ix^j\partial_j-x^ix_j\partial_i\right),\nonumber \\
&& D=x^i\partial_i, ~~~J_{ij}=x_i\partial_j-x_j\partial_i.
\end{eqnarray}
Hence, they generate the boundary conformal group acting on ${\mathbb{R}}^{n}$. 
Representations of the $SO(n+1,1)$ algebra can be constructed by  the method 
of  induced representations \cite{Dixmie,Vilenkin}. The stability group at $x^i=0$ of the $SO(n+1,1)$ algebra is generated by  $(J_{ij},D,K_i)$ and it can easily be verified that $P_i,K_i$ act as 
raising and lowering operators, respectively on  $D$. 
Hence, the  irreducible representation of $SO(n+1,1)$ will be  
states annihilated by $K_i$ and will be specified by their  spin 
and by their conformal dimension $\Delta$, 
\be
&&[K_i,\phi_s(\vec{0})]=0,\nonumber \\
&&[L_{ij},\phi_s(\vec{0})]=\Sigma^{(s)}_{ij}\phi_s(\vec{0}), \nonumber \\
&& [D,\phi_s(\vec{0})]=-i \Delta \phi_s(\vec{0})
,  \label{ls}
\ee
where $\Sigma^{(s)}_{ij}$ is a spin-$s$ representation of $SO(n+1)$. 
The irreducible  representations $\phi_s(\vec{0})$, which are annihilated by $K_i$ and satisfy the relations (\ref{ls}) are the primary fields. If  the primary fields are known, all other fields, 
the descendants,  are constructed by taking derivatives
of the primaries $\partial_i\cdots \partial_j \phi_s(\vec{0})$. 
One of the Casimirs of the $SO(n+1,1)$ algebra is 

\begin{eqnarray}
{\cal C}_1=\frac{1}{2}J_{AB}J^{AB}=D^2+\frac{1}{2}\{P_i,K^i\}+\frac{1}{2} J_{ij} J^{ij}.
\end{eqnarray}
With ${\cal C}_1=m^2/H^2$ we get 
\begin{eqnarray}
\frac{m^2}{H^2}=-\Delta(\Delta-n)-s(s+1),
\end{eqnarray}
  since $\frac{1}{2}\Sigma^{(s)}_{ij}\Sigma^{(s)}_{ij}=s(s+1)$. 
For scalars ($s=0$) in particular,  we have 
\begin{eqnarray}
m^2=-\Delta(\Delta-n)H^2. 
\end{eqnarray}
The conformal dimension of a scalar, in terms of its mass, turns out to be  
\begin{eqnarray}
\label{mainf}
\Delta=\Delta_{\pm}=\frac{n}{2}\left(1\pm \sqrt{1-\frac{4m^2}{n^2H^2}}\right).
\end{eqnarray}
Depending on the mass, the scalar representations of the de Sitter group $SO(n+1,1)$  split into three
distinct series. If the mass is  $ m^2 \geq  n^2H^2/4$, the scalar belongs to the principal series (and can be contracted to representations  of the Poincar\'e algebra), whereas  for a  mass is in the range 
$0\leq m^2\leq n^2H^2/4$, the scalar belongs to the complementary series. Finally, there is also an additional discrete series.

\section{The  two-point function}
Let us now consider a generic  massive scalar field $\phi(\vx,\eta)$ in $\d S_{n+1}$. It may be expanded in spatial Fourier
 modes as 
 \begin{eqnarray}
 \phi(\vx,\eta) =\int \d^n k\, e^{i\vec{k}\cdot \vec{x}}\, a(\vec{k},\eta)+{\rm h.c.}\, .
 \label{phi}
 \end{eqnarray}
Solving the massive scalar equation we find that the Fourier modes $a(\vec{k},\eta)$ are given by
\begin{eqnarray}
•a(\vec{k},\eta)=f_{\vec{k}}(\eta)a(\vec{k}),
\end{eqnarray}
where 
\begin{eqnarray}
f_{\vec{k}}=k^{-\nu}(-\eta)^{n/2}H_\nu(k\eta), ~~~~\nu=\frac{n}{2}\sqrt{1-\frac{4m^2}{n^2H^2}}.
\end{eqnarray}
Hence,
 the scalar field in Eq. (\ref{phi}) can be expressed as \cite{Deser}
\begin{eqnarray}
\phi(\vx,\eta)=\int \d^ny \, K(x,y) \,\Phi(y),
\end{eqnarray}
where $\Phi(y)$ is the scalar profile at $I^+$ and 
\begin{eqnarray}
 K(x,y)=\int \d^n k\, k^{-\nu}\,H_\nu(k\eta) (-\eta)^{n/2}\, e^{i\vec{k}\cdot (\vec{x}-\vec{y})}.
 \end{eqnarray} 
 Defining $\Delta_\pm$ as in Eq. (\ref{mainf}), 
 it turns out that $\Phi(y)$ is a field of conformal weight $\Delta_+$ under the conformal group at $I^+$. 
Then, by using the integral representation of the Hankel function 
\begin{eqnarray}
k^{-\nu}H_\nu(k\eta)=-\frac{i}{\pi}e^{-\frac{i}{2}\nu\pi}\int_0^\infty
\frac{d\tau}{\tau^{\nu+1}}e^{\frac{i}{2}u(\tau+\frac{k^2}{\tau})},
\end{eqnarray}
we find that 
\begin{eqnarray}
K(x,y)=C_\Delta \tilde K_{\Delta}(x,y),
\end{eqnarray}
where 
\begin{eqnarray}
\tilde K_\Delta(x,y)=\left(\frac{\eta}{-\eta^2+|\vec{x}-\vec{y}|^2}\right)^{\Delta}\, , ~~~\Delta=\Delta_-,
\end{eqnarray}
and $C_\Delta$ is a normalization constant. Here 
$\tilde K_\Delta(x,y)$ is just  the bulk to boundary propagator \cite{witten} in the $O^+$ Poincar\'e patch.  
\begin{figure}[H]
    \centering
    \includegraphics[width=0.33\textwidth]{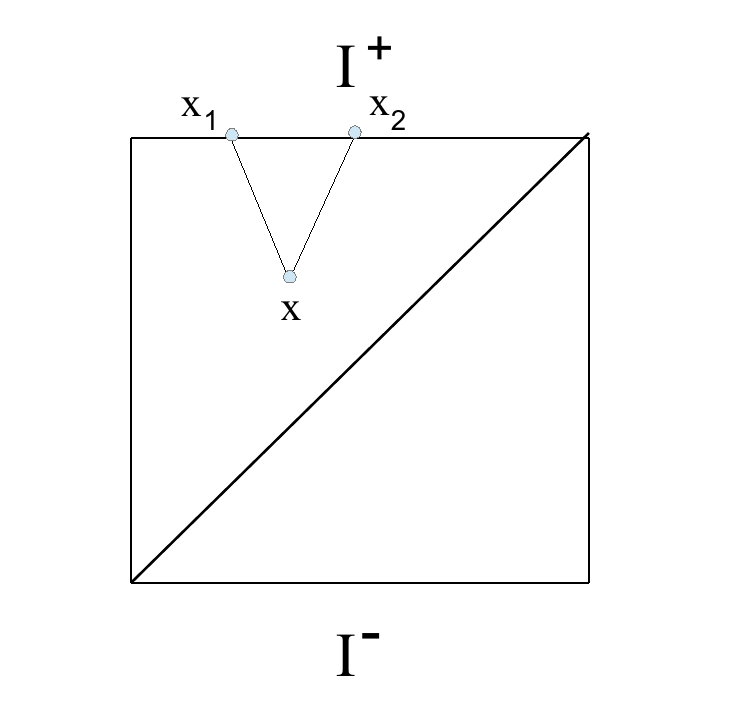}
    \caption{The two-point function for operators at $I^+$.} \label{2pt}
\end{figure}
\noindent
The $N$-point function of  operators ${\cal{O}}_{\Delta_i}$ of dimension $\Delta_i$ in the dual CFT  is generically given  by (up to an overall coefficient) \cite{D'Hoker}
\begin{eqnarray}
\Big< {\cal{O}}_{\Delta_1}(x_1){\cal{O}}_{\Delta_2}(x_2)\cdots{\cal{O}}_{\Delta_N}(x_N)\Big>=
\frac{1}{\pi^{n/2}}\int d^{n+1}x \eta^{-n-1}\prod_{i=1}^N\tilde K_{\Delta_i}(x,x_i).
\end{eqnarray}
Its shape is entirely determined by the three-dimensional conformal symmetry and therefore, as we will explicitly  confirm, any correction has to preserve the shape. 
In particular, for the tree-level two-point function of Fig. \ref{2pt}, we find  
\begin{eqnarray}
\Big< {\cal{O}}_{\Delta}(x_1){\cal{O}}_{\Delta}(x_2)\Big>_0
&=&\frac{1}{\pi^{n/2}}\int \d^{n+1}x \eta^{-n-1}\tilde K_{\Delta}(x,x_1)\tilde K_{\Delta}(x,x_2)\nonumber\\
&=& 
\frac{\Gamma(\Delta)}{4\pi^{3n/2}\Gamma(\Delta-n/2)} \frac{1}{|\vec{x}_1-\vec{x}_2|^{2\Delta}}.
\end{eqnarray}
\subsection{The two-point function in the presence of massive fields}
\noindent
We now consider contributions from exchange of scalar operators of dimension $\Delta_0$ as in Fig. \ref{2pt-e}. This contribution turns out to be
\begin{eqnarray}
\Big< {\cal{O}}_{\Delta}(x_1){\cal{O}}_{\Delta}(x_2)\Big>_{1}
&=&\frac{1}{\pi^{n/2}}\int \frac{\d^{n+1}x}{x_0^{n+1}} \int \frac{\d^{n+1}y}{y_0^{n+1}}\tilde K_{\Delta}(x_1,x)G_{\Delta_0}(x,y)\tilde K_{\Delta}(y,x_2), \label{cor}
\end{eqnarray}
where $G_{\Delta_0}(x,y)$ is the Green function on $\d S_{n+1}$ \cite{Allen,Mottola} and  satisfies the equation
\begin{eqnarray}
 \left(-\nabla_\mu\nabla^\mu +\Delta_0(\Delta_0-n)\right)G_{\Delta_0}(x,y)=x_0^{n+1}
 \delta_D(x-y). \label{green}
 \end{eqnarray} 
\begin{figure}[H]
    \centering
    \includegraphics[width=0.25\textwidth]{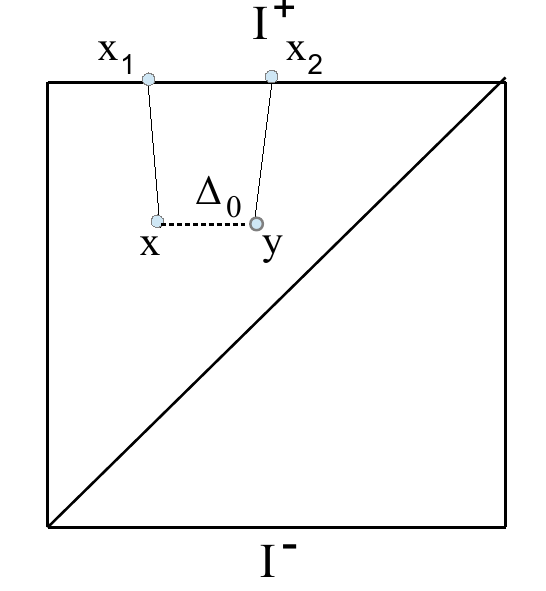}
    \caption{The two-point function for operators at $I^+$ due to exchange of scalar operator in the Poincar\'e patch.} \label{2pt-e}
\end{figure}
\noindent
The Green function $G_{\Delta_0}(x,y)$ which solves (\ref{green}) is given by
\begin{eqnarray}
G_{\Delta_0}(x,y)=\frac{\Gamma(\Delta_{0+})\Gamma(\Delta_{0-})}{(4\pi)^{n/2}\Gamma(n/2)}{}_2F_1\left(\Delta_{0+},\Delta_{0-},\frac{n}{2};\frac{1-u}{2}\right), \label{green}
\end{eqnarray}
where $\Delta_{0\pm}$ are the roots of the equation 
\begin{eqnarray}
m^2=-\Delta_0(\Delta_0-n)H^2
\end{eqnarray}
and 
\begin{eqnarray}
u^2=\frac{x_0^2+y_0^2-|\vec{x}-\vec{y}|^2}{2x_0y_0}
\end{eqnarray}
is the de Sitter invariant length. 
It is easy to verify, by using the analytic properties of the the hypergeometric function in Eq. (\ref{green}), that 
\begin{eqnarray}
G_{\Delta_0}(x,y)\approx y_0 ^{\Delta_{0-}}
\frac{2^{n-1+\Delta_{0-}}}{(2\pi)^{\frac{n-1}{2}}}
\frac{\Gamma(\Delta_{0-})\Gamma(n)}{C_{\Delta_{0-}}}K_{\Delta_{0-}}(x,y)+\cdots,~~~~y_0\to 0.
\end{eqnarray}
The bulk to boundary propagator $K_\Delta (x,y)$ satisfies the equation
\begin{eqnarray}
 \left(-\nabla_\mu\nabla^\mu +\Delta(\Delta-n)\right)\tilde K_{\Delta}(x,y)=0
 \end{eqnarray} 
and the function $J(x,y)$ defined as  
\begin{eqnarray}
J(x,y)=\int \frac{d^{n+1}z}{z_0^{n+1}}G_{\Delta_0}(x,z)K_{\Delta}(z,y)
\end{eqnarray}
obeys the equation
\begin{eqnarray}
\left(-\nabla_\mu\nabla^\mu +\Delta_0(\Delta_0-n)\right)J(x,y)=
\tilde K_{\Delta}(x,y). \label{J}
\end{eqnarray}
The solution to  Eq. (\ref{J}) above is given by 
\begin{eqnarray}
J(x,y)=\frac{1}{\Delta_0(\Delta_0-n)-\Delta(\Delta-n)}\tilde K_{\Delta}(x,y)
\end{eqnarray}
and therefore Eq. (\ref{cor}) is written as 
\begin{eqnarray}
\Big< {\cal{O}}_{\Delta}(x_1){\cal{O}}_{\Delta}(x_2)\Big>_{1}
&=&\frac{1}{\pi^{n/2}}\int \frac{d^{n+1}x}{x_0^{n+1}} \tilde K_{\Delta}(x_1,x)J(x,x_2). \label{cor}
\end{eqnarray}
Note that one should also add the homogeneous solution  
$K_{\Delta_0}(x,x_2)$ to find the general solution to Eq. (\ref{J}), but we do not consider it here as it will 
contribute only for $\Delta_0=\Delta$. 
Then, using the above expression for $J(x,y)$ in Eq. (\ref{cor}) we get  
\begin{eqnarray}
\Big< {\cal{O}}_{\Delta}(x_1){\cal{O}}_{\Delta}(x_2)\Big>_{1}
&=&a\Big< {\cal{O}}_{\Delta}(x_1){\cal{O}}_{\Delta}(x_2)\Big>_0,
\end{eqnarray}
where 
\begin{eqnarray}
a=\frac{1}{\Delta_0(\Delta_0-n)-\Delta(\Delta-n)}.
\end{eqnarray}
For massive scalars we have that 
\begin{eqnarray}
\Delta_0=\Delta_{0\pm}=\frac{n}{2}\left(1\pm \sqrt{1-\frac{4m^2}{n^2H^2}}\right)
\end{eqnarray}
and therefore, for heavy fields $m^2\gg H^2$,  we obtain 
\begin{eqnarray}
\Delta^2_0=-\frac{m^2}{H^2}
 \end{eqnarray}
 Thus,  the corrections to the two-point function due to exchange of heavy scalars with mass $m^2\gg H^2$ is power-law suppressed as 
 \begin{eqnarray}
 a\approx -\frac{H^2}{m^2}.
 \end{eqnarray}
The  two-point function can be written as 
\be
\Big< {\cal{O}}_{\Delta}(x_1){\cal{O}}_{\Delta}(x_2)\Big>=
\Big< {\cal{O}}_{\Delta}(x_1){\cal{O}}_{\Delta}(x_2)\Big>_0+g\Big< {\cal{O}}_{\Delta}(x_1){\cal{O}}_{\Delta}(x_2)\Big>_1+\cdots, 
\ee
or
\be
\fbox{$\displaystyle
\Big< {\cal{O}}_{\Delta}(x_1){\cal{O}}_{\Delta}(x_2)\Big>=
\left(1-g \frac{H^2}{m^2}+\cdots\right)\Big< {\cal{O}}_{\Delta}(x_1){\cal{O}}_{\Delta}(x_2)\Big>_0+\cdots$},
\ee
where $g$ denotes a generic coupling constant and/or a mixing angle (as, for instance, in the case of
quasi-inflation \cite{quasi}). In the case in which we identify the light field with the inflaton field driving inflation, it is easy to see that the heavy fields shift the spectral index $n_s$ of the scalar perturbations by an amount

\be
\Delta n_s=\sum_i 2\, g_i\,\epsilon\,\frac{H^2}{m^2}=\sum_i \, g_i\,\frac{r}{8}\,\frac{H^2}{m^2},
\ee
where $\epsilon=-\dot H/H^2$ is one of the slow-roll parameters and $r=16\epsilon$ (valid for single-field model of inflation) is the tensor-to-scalar ratio. Taking $r\simlt 0.1$, $H/m\simlt 1/5$ and a common coupling of the order of $10^{-1}$, we find $\Delta n_s\simlt 4$ (number of fields)$\,\,\cdot \,\,10^{-4}$, possibly too small to be detected.

There are of course   contributions to the two-point correlators coming from loop corrections as well. 
For example, consider  the Witten diagram of Fig. \ref{2pt-l} below.
\begin{figure}[H]
    \centering
    \includegraphics[width=0.28\textwidth]{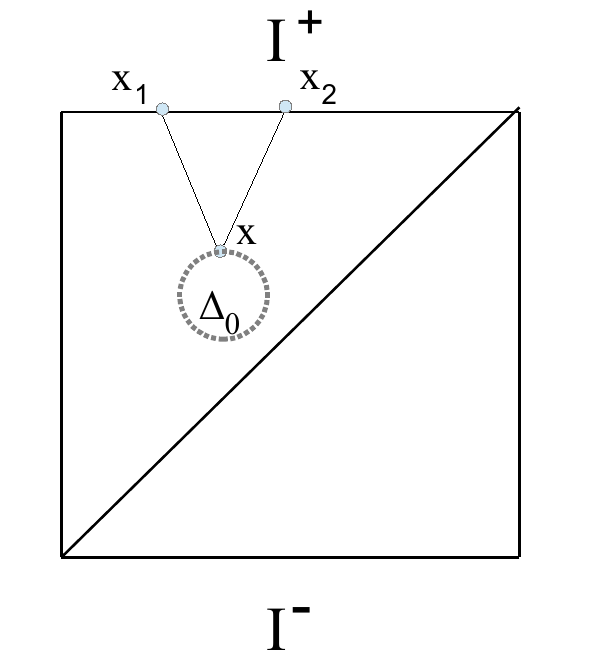}
    \caption{Contribution to the two-point function due to scalar operator
    of dimension $\Delta_0$ running in the loop indicated  at $I^+$.}\label{2pt-l}
\end{figure}
\noindent
Its contribution  to the two-point function is given by 
\begin{eqnarray}
\Big< {\cal{O}}_{\Delta}(x_1){\cal{O}}_{\Delta}(x_2)\Big>_{1L}
&=&\frac{1}{\pi^{n/2}}\int \frac{\d^{n+1}x}{x_0^{n+1}} \tilde K_{\Delta}(x_1,x)G_{\Delta_0}(x,x)\tilde K_{\Delta}(x_2,x). \label{cor1}
\end{eqnarray}
From Eq. (\ref{green}) we get that 
\begin{eqnarray}
G_{\Delta_0}(x,x)=\frac{\Gamma(\Delta_{0+})\Gamma(\Delta_{0-})}{(4\pi)^{n/2}\Gamma(n/2)} ~~~~ (u=1),
\end{eqnarray}
and since for heavy fields we have that 
\begin{eqnarray}
\Delta_{0\pm}\approx\frac{n}{2}\pm i\frac{m}{H}, 
\end{eqnarray}
we get that 
\begin{eqnarray}
G_{\Delta_0}(x,x)\approx \frac{(4\pi)^{\frac{2-n}{2}}}{2\Gamma(n/2)}\left(\frac{m}{H}\right)^{n-1}e^{-\pi m/H}.
\end{eqnarray}
Therefore,
the contribution of Fig. \ref{2pt-l} to the two-point function is given by 
\begin{eqnarray}
\Big< {\cal{O}}_{\Delta}(x_1){\cal{O}}_{\Delta}(x_2)\Big>_{1L}
&=&
\frac{(4\pi)^{\frac{2-n}{2}}}{2\Gamma(\frac{n}{2})}\left(\frac{m}{H}\right)^{n-1}e^{-\pi m/H}
\Big< {\cal{O}}_{\Delta}(x_1){\cal{O}}_{\Delta}(x_2)\Big>_{0}. \label{cor1L}
\end{eqnarray}
As a result, there are power-law suppressions and exponential suppressions as well. 
As we see explicitly, the corrections coming from the heavy fields preserve the shape of the
two-point function dictated by the three-dimensional conformal symmetry. 
\section{The four-point function}
Let us now turn our attention to the four-point function of local operators ${\cal O}_{\Delta_i}$. We have
\begin{eqnarray}
\Big< {\cal{O}}_{\Delta_1}(x_1){\cal{O}}_{\Delta_2}(x_2)
{\cal{O}}_{\Delta_3}(x_3){\cal{O}}_{\Delta_4}(x_4)\Big>_0=
\frac{1}{\pi^{n/2}}\int \d^{n+1}x \eta^{-n-1}
\prod_{i=1}^N\tilde K_{\Delta_i}(x,x_i), \label{4pt}
\end{eqnarray}
as  shown in the Fig. \ref{4pt}.

\begin{figure}[H]
    \centering
    \includegraphics[width=0.28\textwidth]{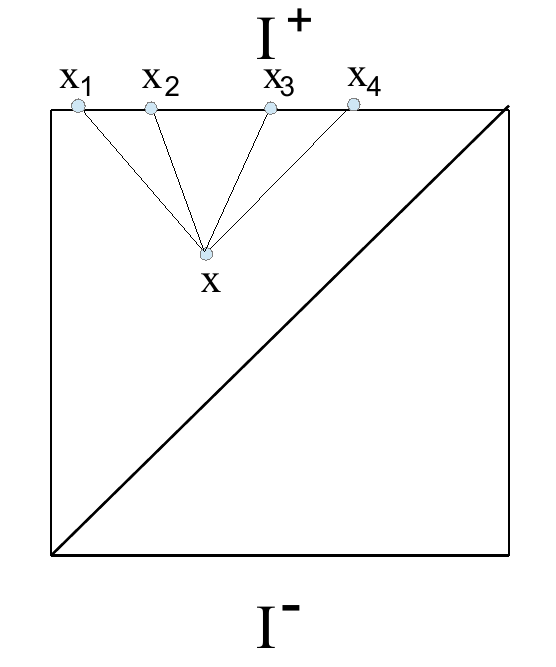}
    \caption{The 4pt-function for operators at $I^+$.} \label{4pt}
\end{figure}
\noindent
 The integrals in Eq. (\ref{4pt})  over the de Sitter space can be done by introducing 
the Schwinger parametrization
\begin{eqnarray}
 \frac{1}{A_1^{a_1}A_2 ^{a_2}\cdots A_n^{a_n}}&=&
 \frac{e^{-i\pi (a_1+a_2+\cdots a_n)}}{\prod_{i=1}^n\Gamma(a_i)}
 \int_0^\infty \d u_1\d u_2\cdots \d u_nu_1^{a_1-1}\cdots u_n^{a_n-1}
 e^{i(u_1A_1+\cdots u_n A_n)}.
 \end{eqnarray} 
We find  that  the $N$-point function in the dual CFT theory  is written as 
\begin{eqnarray}
\Big< {\cal{O}}_{\Delta_1}(x_1)\cdots {\cal{O}}_{\Delta_N}(x_n)\Big>_0=\frac{\pi^{-n/2}e^{-i\pi \Sigma}}{\prod_{i=1}^N\Gamma(\Delta_i)}\int \d^{n}x\,  \d\eta \, \eta^{2 \Sigma -n-1}\int_0^\infty 
\left(\prod_{i=1}^N\d u_i u_i^{\Delta_i-1}\right)e^{i(\sum_{i=1}^Nu_iA_i)},
\end{eqnarray}
where $2\Sigma=\sum_i\Delta_i$ and $A_i$ are given by
\begin{eqnarray}
A_i=-\eta^2+|\vec{x}-\vec{x}_i|^2.\label{A}
\end{eqnarray}
By using Eq. (\ref{A}) and by noticing that 
\begin{eqnarray}
u_1 A_1+u_2 A_2+\cdots u_n A_n=-\eta^2+|\vec{x}|^2+\sum_{i<j}u_{i}u_{j} |\vec{x}_i-\vec{x}_j|^2, \label{AA}
\end{eqnarray}
after appropriate shifts in $\vec{x}$ and  performing the $\eta$ and $\vx$ integrations over de Sitter spacetime, we get
\begin{eqnarray}
 \Big< {\cal{O}}_{\Delta_1}(x_1)\cdots {\cal{O}}_{\Delta_N}(x_n)\Big>_0=\frac{\Gamma(\Sigma-\frac{n}{2})}{2\prod_i\Gamma(\Delta_i)} \int_0^\infty 
\left(\prod_{i=1}^N\d u_i u_i^{\Delta_i-1}\right)\frac{1}{U^\Sigma}e^{\frac{i}{U}\sum_{i<j}u_iu_j x_{ij}^2}, \label{sym}
 \end{eqnarray} 
where 
\begin{eqnarray}
U=\sum_iu_i, ~~~~x_{ij}=|\vec{x}_i-\vec{x}_j|.
\end{eqnarray}
 The integration over the Schwinger parameters ca be done by the Symanzik trick. The latter amounts to replace $\Sigma$  with $u_N$ since the integral does not change (as proven in Ref. \cite{Symanzik}). For example, for the three-point function and choosing 
$S=u_3$, we get after performing the integration over $u_3$ and then over $u_1$ and $u_2$, 
\begin{eqnarray}
 \Big< {\cal{O}}_{\Delta_1}(x_1){\cal{O}}_{\Delta_2}(x_2)
 {\cal{O}}_{\Delta_3}(x_3)\Big>_0\!\!&\!\!=\!\!&\!\!(-i)^{-\Sigma}
 \frac{\Gamma(\Sigma-\frac{n}{2})}{2\Gamma(\Delta_1)\Gamma(\Delta_2)\Gamma(\Delta_3)}\frac{\Gamma(\Sigma-\Delta_1)\Gamma(\Sigma-\Delta_2)\Gamma(\Sigma-\Delta_3)}{(x_{12}^2)^{\Sigma-\Delta_3}(x_{13}^2)^{\Sigma-\Delta_2}(x_{23}^2)^{\Sigma-\Delta_1}}. 
 \end{eqnarray} 
 This is the a well-known result \cite{frad}.

The evaluation of the four-point function is more involved. We may again use Eq. (\ref{sym}) with the $U=u_4$ and   employing the relations \cite{Symanzik}
\begin{eqnarray}
 &&e^{i\frac{u_1u_2}{u_4}x_{12}^2}=\frac{1}{2\pi i}
 \int_{-i\infty}^{+i\infty}\d s\, \Gamma(-s) (-i)^s
 \left(\frac{u_1u_2}{u_4}x_{12}^2\right)^s,\\
 &&e^{i\frac{u_2u_3}{u_4}x_{23}^2}=\frac{1}{2\pi i}
 \int_{-i\infty}^{+i\infty}\d t\, \Gamma(-t) (-i)^t
 \left(\frac{u_2u_3}{u_4}x_{23}^2\right)^t.
 \end{eqnarray} 
Then, integrating $u_4$ and then $u_1$, $u_2$ and $u_3$ we find that 
\begin{eqnarray}
\Big< {\cal{O}}_{\Delta_1}(x_1){\cal{O}}_{\Delta_2}(x_2)
{\cal{O}}_{\Delta_3}(x_3){\cal{O}}_{\Delta_4}(x_4)\Big>_0&=& \frac{\Gamma(\Sigma-\frac{n}{2})}{2\prod_i\Gamma(\Delta_i)}(-i)^{-\Sigma}
\label{q1} \\
&\times& (x_{14}^2)^{\Sigma-\D_1-\D_4}(x_{34}^2)^{\Sigma-\D_3-\D_4}(x_{13}^2)^{\D_4-\Sigma}(x_{24}^2)^{-\D_2}\nonumber \\
&\times& G(\D_2,\Sigma-\D_4,\D_3+\D_4-\Sigma,\D_1+\D_4-\Sigma|u,v),\nonumber
\end{eqnarray}
where 
\begin{eqnarray}
u=\frac{x_{12}^2x_{34}^2}{x_{13}^2x_{24}^2}, ~~~
v=\frac{x_{14}^2x_{23}^2}{x_{13}^2x_{24}^2},
\end{eqnarray}
and 
\begin{eqnarray}
&&\hspace{-2.2cm}G(\D_2,\Sigma-\D_4,\D_3+\D_4-\Sigma,\D_1+\D_4-\Sigma|u,v)=
\nonumber \\
\hspace{2cm}&-&\frac{1}{(2\pi)^2}\int_{-i\infty}^{+i\infty} \d s\int_{-i\infty}^{+i\infty}\d t\, \Gamma(-s)\Gamma(-t)\Gamma(\D_2+s+t)
\Gamma(\Sigma-\D_4+s+t)\nonumber \\\hspace{2cm}&\times& 
\Gamma(\D_3+\D_4-\Sigma-s) \Gamma(\D_1+\D_4-\Sigma-t)\, u^sv^t. \label{st}
\end{eqnarray}
The integrals in Eq. (\ref{st}) can be done in the complex plane by choosing a contour for ${\rm Re}\,s>0$ and ${\rm Re} \, t>0$. The result is \cite{DT}
\begin{eqnarray}
&&\hspace{-2.2cm}G(\D_2,\Sigma-\D_4,\D_3+\D_4-\Sigma,\D_1+\D_4-\Sigma|u,v)=
\nonumber \\
&&\Gamma(\D_2)\Gamma(\Sigma-\D_4)\Gamma(\D_3+\D_4-\Sigma)\Gamma(\D_1+\D_4-\Sigma)\nonumber \\
&&\times F_4(\D_2,\Sigma-\D_4,\Sigma+1-\D_3-\D_4,\Sigma+1-\D_1-\D_4;u,v)
\nonumber \\
&&+\Gamma(\Sigma-\D_3)\Gamma(\D_1)\Gamma(\D_3+\D_4-\Sigma)\Gamma(\D_2+\D_3-\Sigma)\nonumber \\
&&\times v^{\D_1+\D_2-\Sigma}F_4(\Sigma-\D_3,\D_1,\Sigma+1-\D_3-\D_4,\Sigma+1-\D_2-\D_3;u,v) \label{g}
\\ && +
\Gamma(\Sigma-\D_1)\Gamma(\D_3)\Gamma(\D_1+\D_2-\Sigma)\Gamma(\D_1+\D_4-\Sigma)u^{\D_3+\D_4-\Sigma}\nonumber \\
&&
\times F_4(\Sigma-\D_1,\D_3,\Sigma+1-\D_1-\D_2,\Sigma+1-\D_3-\D_4;u,v)
\nonumber \\
&&
+ \Gamma(\D_4)\Gamma(\Sigma-\D_2)\Gamma(\D_1+\D_2-\Sigma)\Gamma(\D_2+\D_3-\Sigma)\nonumber \\
&& \times u^{\D_3+\D_4-\Sigma}v^{\D_1+\D_4-\Sigma}
F_4(\D_4,\Sigma-\D_2,\Sigma+1-\D_1-\D_2,\Sigma+1-\D_2-\D_3;u,v),\nonumber 
\end{eqnarray}
where \cite{erdelyi}
\begin{eqnarray}
F_4(a,b,c,d;x,y)=\sum_{j=0}^{\infty}\sum_{\ell=0}^{\infty}
\frac{(a)_{j+\ell}(b)_{j+\ell}}{(c)_{j}(d)_\ell}\frac{x^jy^\ell}{j!\ell!} 
\end{eqnarray}
 is Appel's hypergeometric function of two variables and  $(a)_j=\Gamma(a+j)/\Gamma(a)$ is the Pochhammer symbol. The appearance of Appel's hypergeometric functions in the four-point function has been noticed in Ref. \cite{FGG}. 
It is important to note here that the radius of convergence of 
$F_4(a,b,c,d;x,y)$ is 
\begin{eqnarray}
\sqrt{x}+\sqrt{y}<1. \label{xy}
\end{eqnarray}
As a result, the Appel's functions appearing in Eq. (\ref{g}) converges for
\begin{eqnarray}
\sqrt{u}+\sqrt{v}<1,
\end{eqnarray}
or
\begin{eqnarray}
x_{13}x_{24}>x_{12}x_{34}+x_{14}x_{23}\,. \label{conv}
\end{eqnarray}
However, from Ptolemy's inequality we know  that for any quadrilateral $ABCD$ we have
\begin{eqnarray}
\overline{AC} \cdot  \overline{BD}<\overline{AB}\cdot \overline{CD} +\overline{BC}\cdot \overline{AD}. \label{ptolemy}
\end{eqnarray}
Taking the points $A,B,C,D$ to be at positions $\vec{x}_1,\vec{x}_2,\vec{x}_3$ and $\vec{x}_4$ respectively, we find that 
\begin{eqnarray}
x_{13}x_{24}<x_{12}x_{34}+x_{14}x_{23}\,. 
\end{eqnarray}
and therefore Eq. (\ref{conv}) violates Ptolemy's inequality. Therefore, Appel's $F_4$ should be extended beyond its convergence region. This issue is discussed in Appendix A.
\subsection{The four-point function in the presence of massive fields}
\noindent
Let us now consider the corrections to the four-point function due to scalar exchange as shown in Fig. \ref{4pt-e}. 
\begin{figure}[H]
    \centering
    \includegraphics[width=0.28\textwidth]{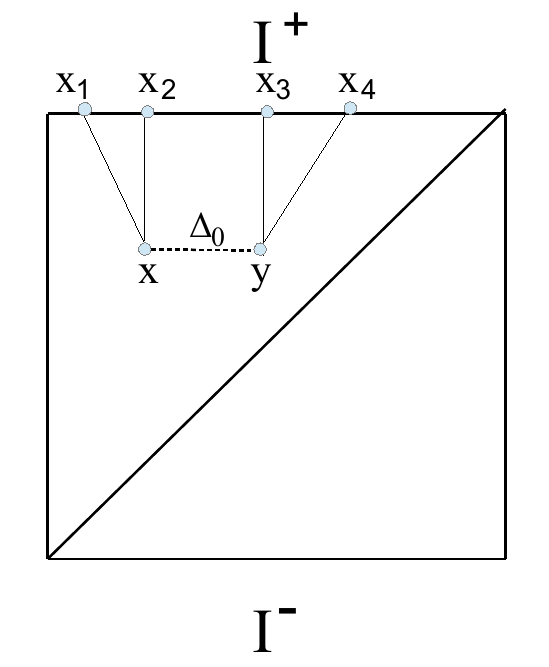}
    \caption{The four point function of operators at $I^+$ due to exchange of scalar operator in the Poincar\'e patch.} \label{4pt-e}
\end{figure}
\noindent
The contribution to the four-point function by the exchange of a scalar operator of dimension $\Delta_0$ is  given by 
\begin{eqnarray}
\Big< {\cal{O}}_{\Delta_1}(x_1){\cal{O}}_{\Delta_2}(x_2)
{\cal{O}}_{\Delta_23}(x_3){\cal{O}}_{\Delta_4}(x_4)\Big>_{1}
&=&\frac{1}{\pi^{n/2}}\int \frac{\d^{n+1}x}{x_0^{n+1}} \int \frac{\d^{n+1}y}{y_0^{n+1}}\tilde K_{\Delta_1}(x,x_1)\tilde K_{\Delta_2}(x,x_2)\nonumber \\
&\times&
G_{\Delta_0}(x,y)\tilde K_{\Delta_3}(y,x_3)\tilde K_{\Delta_4}(y,x_4), \label{cor4}
\end{eqnarray}
where again $G_{\Delta_0}(x,y)$ is the Green function on $\d S_{n+1}$ given in Eq. (\ref{green}).
It is easy then to see that 
\begin{eqnarray}
J(x,x_3,x_4)=\int \frac{\d^{n+1}y}{y_0^{n+1}}G_{\Delta_0}(x,y)\tilde K_{\Delta_3}(y,x_3)\tilde K_{\Delta_4}(y,x_4)\label{j4}
\end{eqnarray}
satisfies the equation 
\begin{eqnarray}
(-\nabla^2+\D_0(\D_0-n))J(x,x_3,x_4)=\tilde K_{\Delta_3}(x,x_3)\tilde K_{\Delta_4}(x,x_4).  \label{eq4}
\end{eqnarray}
The solution to this equation has been worked out in Refs. \cite{Dolan1,Dolan2} for the anti-de Sitter background, but it can easily be worked out for the de Sitter case as well. The result is 
\begin{eqnarray}
J(x,x_3,x_4)&=&-\frac{1}{4} \sum_{m=0}
\frac{(\D_3)_s(\D_4)_m \,x_{34}^{2m}}{(\frac{1}{2}(\D_3+\D_4-\D_0))_{m+1}(\frac{1}{2}
(\D_3+\D_4+\D_0-n))_{m+1}} \nonumber \\
&\times& \tilde K_{\Delta_3+m}(y,x_3)\tilde K_{\Delta_4+m}(y,x_4)\nonumber \\
& +& A\,  \sum_{m=0} \frac{(\frac{1}{2}(\D_0+\D_3-\D_4))_m(\frac{1}{2}(\D_0+\D_4-\D_3))_m\, x_{34}^{\D_0-\D_3-\D_4+2m}}{m!(\D_0-\frac{n}{2}+1)_m}\nonumber \\
&\times& 
\tilde K_{\frac{1}{2}(\D_0+\Delta_3-\D_4)+m}(y,x_3)\tilde K_{\frac{1}{2}(\D_0+\Delta_4-\D_3)+m}(y,x_4), \label{J4}
\end{eqnarray}
where 
\begin{eqnarray}
A&=&\frac{1}{4} \Gamma\left(\frac{1}{2}(\D_3+\D_4+\D_0-n)\right)
\frac{\Gamma\left(\frac{1}{2}(\D_3+\D_4-\D_0)\right)\Gamma\left(\frac{1}{2}(\D_0+\D_3-\D_4)\right)\Gamma\left((\frac{1}{2}(\D_0+\D_4-\D_3)\right)}{
\Gamma\left(\D_0-\frac{n}{2}+1\right)\Gamma\D_3)\Gamma(\D_4)}. \nonumber\\
&&
\end{eqnarray}
In particular, the solution  is the sum of two terms. The first one is a partial solution of the inhomogeneous Eq. (\ref{eq4}), whereas the second part is the solution to its  homogeneous counterpart.  
Using Eq. (\ref{J4}) in Eq. (\ref{cor4}) and after performing the $\vx$-integration, we get that the the correction to the four-point function due to scalar exchange of scaling dimension $\D_0$ is
\begin{eqnarray}
\Big< {\cal{O}}_{\Delta_1}(x_1){\cal{O}}_{\Delta_2}(x_2)
{\cal{O}}_{\Delta_23}(x_3){\cal{O}}_{\Delta_4}(x_4)\Big>_{1}
&=&\frac{(-i)^{-\Sigma}}{8\prod_i\Gamma(\Delta_i)}
\frac{(x_{14}^2)^{\Sigma-\D_1-\D_4}(x_{34}^2)^{\Sigma-\D_3-\D_4}}{(x_{13}^2)^{\Sigma-\D_4}(x_{24}^2)^{\D_2}}
D(u,v),
\end{eqnarray}
where $D(u,v)$ is given by
\begin{eqnarray}
D(u,v)=D_{\rm nh}(u,v)+D_{\rm h}(u,v),
\end{eqnarray}
with
\begin{eqnarray}
D_{\rm nh}(u,v)&=&-\sum_{m=0}\frac{\Gamma(\Sigma-\frac{n}{2}-m)}{(\frac{1}{2}(\D_3+\D_4-\D_0))_{m+1}(\frac{1}{2}
(\D_3+\D_4+\D_0-n))_{m+1}} \nonumber \\
&& \times G(\D_2,\Sigma-\D_4,\D_3+\D_4-\Sigma-m,\D_1+\D_4-\Sigma|u,v)
\end{eqnarray}
and 
\begin{eqnarray}
D_{\rm h}(u,v)&=& \frac{\Gamma(\frac{1}{2}(\D_3+\D_4+\D-n)\Gamma(\frac{1}{2}(\D_3+\D_4-\D_0)}{\Gamma(\D_0+\frac{n}{2}+1)}\nonumber \\
&&
\times \sum_{m=0}\frac{ \Gamma\left(\frac{1}{2}(\D_1+\D_2+\D_0-n)+m\right)}{m! (\D_0-\frac{n}{2}+1)_m}\nonumber \\
&&\times G\left(\D_2,\Sigma-\D_4,\frac{1}{2}(\D_3+\D_4-\D_0)-m,\D_1+\D_4-\Sigma|u,v\right).
\end{eqnarray}
We see that there are two contributions to $D(u,v)$, the non-homogeneous contribution from $D_{\rm nh}(u,v)$ and the homogeneous one $D_{\rm h}(u,v)$. 
It is easy to see that the non-homogeneous piece of the scalar exchange contribution is suppressed as 
\begin{eqnarray}
 D_{\rm nh}(u,v)\approx -\frac{4\Gamma(\Sigma-\frac{n}{2})}{\D_0^2}G
 +\cdots
 \end{eqnarray} 
 where 
 \begin{eqnarray}
G&=& G(\D_2,\Sigma-\D_4,\D_3+\D_4-\Sigma,\D_1+\D_4-\Sigma|u,v).
 \end{eqnarray}
Hence, for heavy scalars with $m\gg n H/2$ we have that $\Delta^2_0\approx -m^2/H^2$ so that 
\begin{eqnarray}
 D_{\rm nh}(u,v)\approx 4\, \frac{H^2}{m^2}\, \, \Gamma\left(\Sigma-\frac{n}{2}\right) G+\cdots.
 \end{eqnarray}
Similarly, the suppression of the homogeneous part $D_{\rm h}(u,v)$ can be found by recalling 
that
\begin{eqnarray}
&&\hspace{-2.2cm}G\left(\D_2,\Sigma-\D_4,\frac{1}{2}(\D_3+\D_4-\D_0)-m,\D_1+\D_4-\Sigma|u,v\right)=
\nonumber \\
\hspace{1.5cm}&-&\frac{1}{(2\pi)^2}\int_{-i\infty}^{+i\infty} \d s\int_{-i\infty}^{+i\infty}\d t\, \Gamma(-s)\Gamma(-t)\Gamma(\D_2+s+t)
\Gamma(\Sigma-\D_4+s+t)\nonumber \\\hspace{1.5cm}&\times& 
\Gamma\left(\frac{1}{2}(\D_3+\D_4-\D_0)-m-s\right) \Gamma(\D_1+\D_4-\Sigma-t)\, u^sv^t. \label{st1}
\end{eqnarray}
Then by using 
\begin{eqnarray}
\frac{\Gamma(\D_0+a)}{\Gamma(\D_0)}\approx \D_0^a+\cdots, 
\end{eqnarray}
we get that 
\begin{eqnarray}
G\left(\D_2,\Sigma-\D_4,\frac{1}{2}(\D_3+\D_4-\D_0)-m,\D_1+\D_4-\Sigma|u,v\right)\approx
\Gamma\left(-\frac{\D_0}{2}\right)G_n+\cdots,
\end{eqnarray}
where $G_n$ has a power law dependence on $\Delta_0$ and 
%
we find that 
\begin{eqnarray}
D_{\rm h}(u,v)&=&\frac{\Gamma(\frac{1}{2}(\D_3+\D_4+\D_0-n)\Gamma(\frac{1}{2}(\D_3+\D_4-\D_0)}{\Gamma(\D_0+\frac{n}{2}+1)}\Gamma\left(\frac{1}{2}(\D_1+\D_2+\D_0-n\right)\Gamma\left(-\frac{\D_0}{2}\right) G_n+\cdots. \nonumber\\
&&
\end{eqnarray}
Hence, for heavy scalars
 with $m\gg n H/2$ we have that $\Delta^2\approx -m^2/H^2$ and we find 
 \begin{eqnarray}
  D_{\rm h}(u,v)\approx e^{-\pi m/H}\, \, \frac{(-1)^{\D_3+\D_4-\frac{1}{4}}}{2^{\D_0-\frac{n+1}{2}}}\left(-\frac{m^2}{H^2}\right)^{\Sigma+\D_3+\D_4-\D_0-2}\left(\frac{H}{m}\right)^{i\frac{H}{m}}\, G_n+\cdots.
  \end{eqnarray} 
As a result, the contribution of a heavy scalar exchange $(m>n H/2)$ 
to the four-point function has two contributions. The first one is suppressed by $H^2/m^2$ whereas the second one is exponentially suppressed by the Boltzmann factor $e^{-\pi m/H}$. As a result, the overall suppression for the four-point function due to heavy scalar fields is dominated by $D_{\rm nh}$ 
and we have 

\begin{eqnarray}
\Big< {\cal{O}}_{\Delta_1}(x_1){\cal{O}}_{\Delta_2}(x_2)
{\cal{O}}_{\Delta_23}(x_3){\cal{O}}_{\Delta_4}(x_4)\Big>_{1}
&\approx&\frac{(-i)^{-\Sigma}}{8\prod_i\Gamma(\Delta_i)}
\frac{(x_{14}^2)^{\Sigma-\D_1-\D_4}(x_{34}^2)^{\Sigma-\D_3-\D_4}}{(x_{13}^2)^{\Sigma-\D_4}(x_{24}^2)^{\D_2}}
D_{\rm nh}(u,v)\nonumber \\
&&\hspace{-1cm}=
\frac{(-i)^{-\Sigma}}{2\prod_i\Gamma(\Delta_i)}
\frac{(x_{14}^2)^{\Sigma-\D_1-\D_4}(x_{34}^2)^{\Sigma-\D_3-\D_4}}{(x_{13}^2)^{\Sigma-\D_4}(x_{24}^2)^{\D_2}}
\frac{H^2}{m^2}\, \, \Gamma\left(\Sigma-\frac{n}{2}\right) \nonumber \\
&&\hspace{-1cm} \times 
 G(\D_2,\Sigma-\D_4,\D_3+\D_4-\Sigma,\D_1+\D_4-\Sigma|u,v)+\ldots
 \label{q2}
\end{eqnarray}
Comparing Eqs. (\ref{q1}) and (\ref{q2}), we see that 
\begin{eqnarray}
\Big< {\cal{O}}_{\Delta_1}(x_1){\cal{O}}_{\Delta_2}(x_2)
{\cal{O}}_{\Delta_23}(x_3){\cal{O}}_{\Delta_4}(x_4)\Big>_{1}= 
\frac{H^2}{m^2}\Big< {\cal{O}}_{\Delta_1}(x_1){\cal{O}}_{\Delta_2}(x_2)
{\cal{O}}_{\Delta_23}(x_3){\cal{O}}_{\Delta_4}(x_4)\Big>_{0}+\ldots
\end{eqnarray}
and therefore, 

$$
\fbox{$\displaystyle
\Big< {\cal{O}}_{\Delta_1}(x_1){\cal{O}}_{\Delta_2}(x_2){\cal{O}}_{\Delta_3}(x_3){\cal{O}}_{\Delta_4}(x_4)\Big>=
\left(1+g^2 \frac{H^2}{m^2}+\cdots\right)\Big< {\cal{O}}_{\Delta_1}(x_1){\cal{O}}_{\Delta_2}(x_2){\cal{O}}_{\Delta_3}(x_3){\cal{O}}_{\Delta_4}(x_4)\Big>_0+\cdots$}.
$$
\be
\ee
where we have also inserted a coupling $g$ for each vertex of Fig. \ref{4pt-e}. 

We should point out here that, for simplicity, we have considered above only scalar operators. Our results can easily be generalized in the case of operators in non-trivial representations of the $SO(n)$ subgroup of the
 full $SO(n+1,1)$ isometry group or derivative interactions like for example $\phi_1\nabla\phi_2 \nabla\phi_3$. Indeed, by integration by parts and field redefinition one can always reduce the interaction to a cubic one. There is always a power-law suppression even when higher-spin operators in general are involved, due to exchange of heavy scalars.

\section{Conclusions}
\noindent
In this paper we have investigated the impact of heavy fields on the light fields during the inflationary period by making use of the 
dS/CFT correspondance. 
Heavy and light fields belong to different 
representations of the de Sitter group $SO(1, 4)$. 
In particular, heavy fields belong to the principal representation and correspond
to operators with imaginary dimensions in the dual CFT. Light fields, on the other hand, belong to the complementary series. The fact that
heavy fields appear with imaginary dimensions 
leads to a non-unitary theory. However, this is not in contradiction with any basic principle as there is no, strictly speaking, $S$-matrix in de Sitter space and no obvious reason for the unitarity of the dual CFT.

We have first studied the two-point correlator and  found that tree-level corrections due to exchange of bulk heavy fields are power-law suppressed, whereas loop corrections  give an exponential suppression. Similar conclusions have been drawn for the four-point correlator. Already at the tree level the exchange of heavy fields leads to both power-law and exponential suppression.  Our results here indicate that,  beyond 
the expected  Boltzmann suppression understandable from  the thermal properties of de Sitter space, the contribution of the heavy fields to the light field correlators is power-law suppressed. 

Let us stress again that, as we mentioned in the introduction,  our results have been obtained 
assuming that the Hubble rate is the only relevant time scale:   the there is no  breaking of adiabaticity in the inflationary dynamics or a correspondent breaking of three-dimensional conformal symmetry in the dual CFT.

\section*{Acknowledgments}

 The research of A.K. was implemented under the Aristeia II Action of the Operational Programme Education and
Lifelong Learning and is co-funded by the European Social Fund (ESF) and National Resources. A.K. is also partially
supported by European Union’s Seventh Framework Programme (FP7/2007-2013) under REA grant agreement no.
329083. A.R. is supported by the Swiss National Science Foundation (SNSF), project ‘The non-Gaussian Universe”
(project number: 200021140236).


\renewcommand{\theequation}{A.\arabic{equation}}
\setcounter{equation}{0}
\section*{Appendix}

 The  function $F_4$  outside the convergence region (\ref{xy}) is not well-known basically because of its limited known  analytical continuation properties. For example, one might use the relation 

\begin{eqnarray}
F_4(a,b;c,c';u,v)&=&\frac{\Gamma(c')\Gamma(b-a)}{\Gamma(c'-a)\Gamma(b)}\left(e^{\pi}v\right)^{-a}F_4\left(a,a+1-c';c,a+1-b;\frac{u}{v},\frac{1}{v}\right)\nonumber \\
&&+\frac{\Gamma(c')\Gamma(a-b)}{\Gamma(c'-b)\Gamma(a)}\left(e^{\pi}v\right)^{b}F_4\left(b+1-c',b,c,b+1-a;\frac{u}{v},\frac{1}{v}\right).
\end{eqnarray}
However, this  does not help. Although the radius of convergence is now 
\begin{eqnarray}
\sqrt{u}+1<\sqrt{v},
\end{eqnarray}
it still violates Ptolemy's inequality (\ref{ptolemy}). 

The best one can do is to use reduction formulas for Appel's $F_4$ which will allow to continue it in the region $x_{13}x_{24}<x_{12}x_{34}+x_{14}x_{23}$. Below we list all known factorizations of $F_4$ (to the best of our knowledge) where some addition reductions obtained by using contiguity relations are ignored:
\begin{enumerate}
\item For $a+b+1=c+c'$ we have 
\begin{eqnarray}
F_4(a,b,c,c';x(1-y),y(1-x))={}_2F_1(a,b,c,x){}_2F_1(a,b,c',y).\label{n1}
\end{eqnarray}
\item For $c=c'=b$, we have  
\begin{eqnarray}
F_4\left(a,b,b,b;-\frac{x}{(1-x)(1-y)},-\frac{y}{(1-x)(1-y)}\right)
&=&(1-x)^a(1-y)^b\nonumber\\ &&\times\, \,   {}_2F_1(a,a+1-b,b;xy).
\label{n2}
\end{eqnarray}
\item For $c=1+a-b$ and $c'=b$ we have  
\begin{eqnarray}
F_4\left(a,b,1+a-b,b;-\frac{x}{(1-x)(1-y)},-\frac{y}{(1-x)(1-y)}\right)
&=&(1-y)^a\nonumber\\ &&\hspace{-4cm}\times  {}_2F_1\left(a,a+1-b,b;-\frac{x(1-y)}{1-x}\right).
\end{eqnarray}
\item For $c=a$ and $c'=b$ we have
\begin{eqnarray}
F_4\left(a,b,a,b;-\frac{x}{(1-x)(1-y)},-\frac{y}{(1-x)(1-y)}\right)
=\frac{(1-x)^b(1-y)^a}{1-xy}.
\end{eqnarray}
\item For $c'=c$ we have 
\begin{eqnarray}
F_4\left(a,b,c,b;-\frac{x}{(1-x)(1-y)},-\frac{y}{(1-x)(1-y)}\right)
&=&(1-x)^a(1-y)^b\nonumber\\ &&\hspace{-2cm}\times\, \,  F_1(a,c-b,1+a-c,c;x,xy).
\end{eqnarray}
\item For $b=a+\frac{1}{2}$ and $c'=\frac{1}{2}$ we have
\begin{eqnarray}
F_4\left(a,a+\frac{1}{2},c,\frac{1}{2};x y\right)&=&\frac{1}{2}(1+\sqrt{y})^{-2a}{}_2F_1\left(a,a+\frac{1}{2},c;\frac{x}{(1+\sqrt{y})^2}\right)\nonumber \\&&
+\frac{1}{2}(1-\sqrt{y})^{-2a}{}_2F_1\left(a,a+\frac{1}{2},c;\frac{x}{(1-\sqrt{y})^2}\right).
\end{eqnarray}
\end{enumerate}
For example, we may use the reduction (\ref{n1}) when $\Sigma=-1$. In this case all Appel's $F_4(a,b,c,c';u,v)$ functions appearing in the expression (\ref{g}) satisfy $a+b+1=c+c'$ and therefore they factorize as in Eq. (\ref{n1}). In particular, defining \cite{GN}
\begin{eqnarray}
\alpha=x_{12}x_{34}, ~~~\beta=x_{14} x_{23},~~~\gamma=x_{13} x_{24}
\end{eqnarray}
and 
\begin{eqnarray}
x=\frac{\alpha}{\gamma}e^{-i\theta_x}, ~~~y=\frac{\beta}{\gamma}e^{-i\theta_y},
\end{eqnarray}
where 
\begin{eqnarray}
&&\cos\theta_x=\frac{\alpha^2+\gamma^2-\beta^2}{2\alpha \gamma},~~~\sin\theta_x=\frac{\sqrt{\delta}}{2\alpha\gamma}\, , ~~~\delta=[(a+b)^2-c^2][(a-b)^2-c^2],\\
&&\cos\theta_y=\frac{\beta^2+\gamma^2-\alpha^2}{2\beta\gamma},~~~\sin\theta_y=\frac{\sqrt{\delta}}{2\beta\gamma},
\end{eqnarray}
we have 
\begin{eqnarray}
F_4(a,b,c,c';u,v)&=&F_4(a,b,c,c';x(1-y),y(1-x))\nonumber \\
&=&{2}F_1(a,b,c;\sqrt{u}e^{-i\theta_x})
{2}F_1(a,b,c;\sqrt{v}e^{-i\theta_y}).
\end{eqnarray}
Then, it is easy to verify that the function $G$ defined in Eq. (\ref{g}) is convergent for $\D_i+\D_j<1$ (for $i,j=1,...,4)$.

\end{document}